# Energy surface and lifetime of magnetic skyrmions

Uzdin V.M.[1,2], Potkina M.N.[2,3], Lobanov I.S.[1], Bessarab P.F.[1,3], Jónsson H.[3,4]

[1] University ITMO, St. Petersburg, Russia
[2] SPbGU, St. Petersburg, Russia
[3] University of Iceland, Reykjavík, Iceland
[4] Dept. of Applied Physics, Aalto University, Finland

**Abstract**

The stability of skyrmions in various environments is estimated by analyzing the multidimensional surface describing the energy of the system as a function of the directions of the magnetic moments in the system. The energy is given by a Heisenberg-like Hamiltonian that includes Dzyaloshinskii-Moriya interaction, anisotropy and external magnetic field. Local minima on this surface correspond to the ferromagnetic and skyrmion states. Minimum energy paths (MEP) between the minima are calculated using the geodesic nudged elastic band method. The maximum energy along an MEP corresponds to a first order saddle point on the energy surface and gives an estimate of the activation energy for the magnetic transition, such as creation and annihilation of a skyrmion. The pre-exponential factor in the Arrhenius law for the rate, the so-called attempt frequency, is estimated within harmonic transition state theory where the eigenvalues of the Hessian at the saddle point and the local minima are used to characterize the shape of the energy surface. For some degrees of freedom, so-called "zero modes", the energy of the system remains invariant. They need to be treated separately and give rise to temperature dependence of the attempt frequency. As an example application of this general theory, the lifetime of a skyrmion in a track of finite width for a PdFe overlayer on a Ir(111) substrate is calculated as a function of track width and external magnetic field. Also, the effect of non-magnetic impurities is studied. Various MEPs for annihilation inside a track, via the boundary of a track and at an impurity are presented. The attempt frequency as well as the activation energy has been calculated for each mechanism to estimate the transition rate as a function of temperature.

## 1. Introduction

A quantitative measure of the topological stabilization of skyrmion states is an important fundamental problem connecting topology and physics [1]. For a continuum, the difference in topological charge of the homogeneous ferromagnetic (FM) and skyrmion states makes skyrmions absolutely stable with respect to perturbations such as thermal fluctuations. In real systems magnetic moments are localized on sites of a discrete lattice and topological arguments are strictly speaking not valid. Instead the states are separated by a finite activation barrier which becomes infinite only in the continuous limit.

The stability of magnetic skyrmions with respect to thermal fluctuations is a principal issue for their application in devices such as magnetic data storage [2]. Stochastic modeling of skyrmion dynamics to assess stability proves to be a difficult problem due to the large difference in time scales between a typical vibrational period of magnetic moments and the lifetime of a reasonably stable skyrmion. Nevertheless, it is this timescale difference that makes it possible to apply statistical approaches such as transition state theory (TST) to calculate the lifetime of skyrmions as a function of temperature. An Arrhenius law can be obtained for the rates of magnetic transitions and within the framework of harmonic TST (HTST) both the activation energy as well as the attempt frequency can be calculated with controlled accuracy [3,4]. Analysis of thermal stability within HTST is based on an analysis of the multidimensional energy surface of the system, namely the identification of the local minima corresponding to stable states and location of minimum energy path (MEP) between them. Having the largest statistical weight among neighbor paths, the MEP represents the mechanism of the magnetic transition, i.e. most probable magnetic configurations during the transition. The basic TST approach based on the following approximations: (i) classical dynamics of transition, i.e. the possibility of quantum mechanical tunneling, which becomes dominant mechanism at low enough temperature, is not taken into account [5]; (ii) the lifetime is long with respect to spin vibrations so Boltzmann distribution is established and maintained; (iii) trajectories that make it to a transition state dividing surface separating initial and final states and are at heading from there away from the initial state are assumed to end up in the final state, i.e. the transition state is only crossed once. HTST approximation additionally presupposes that the energy surface in the relevant region around the minima and saddle points can be described by a quadratic function for all degrees of freedom. In this case the rate of magnetic transitions can be calculated analytically based on the shape of the energy surface near the minima and saddle point.

A minimum energy path for skyrmion annihilation has been calculated using the GNEB method by several workers [6-10]. The activation energy obtained from the first order saddle point in this way has been found to be in close agreement with Landau-Lifschitz-Gilbert simulations for marginally stable skyrmions where such simulations can be carried out for a sufficiently long time interval [7,11], and HTST is expected to work even better for conditions where the skyrmion is more stable.

## 2. Model and method

The energy surface of the system is described by a Heisenberg type Hamiltonian for spins arranged on a planar triangular lattice. In addition to the exchange interaction it includes out-of-plane magnetic anisotropy, Dzyaloshinsky-Moriya interaction and interaction with an external magnetic field

$$H = -J \sum_{<i,j>} \mathbf{S}_i \mathbf{S}_j - K \sum_i S_{i.z}^2 - \sum_{<i,j>} \mathbf{D}_{ij} [\mathbf{S}_i \times \mathbf{S}_j] - \mu \sum_i \mathbf{B} \, \mathbf{S}_i \qquad (1)$$

Here, $S_i$ is the unit vectors specifying the direction of the magnetic moment on site $i$, $\mu = 3 \, \mu_B$ is the value of magnetic moment which is taken to be the same for all sites. Parameters $J$ and $K$, and vector $\mathbf{D}_{ij}$ specify the magnitude of the exchange, anisotropy and Dzyaloshinsky-Moriya interactions, resp. $\mathbf{D}_{ij}$ is chosen to lie the plane of the lattice perpendicular to the vector connecting atomic sites $i$ and $j$. The magnetic field, $\mathbf{B}$, is taken to be perpendicular to the plane, i.e. points along the z-axis of the coordinate system. The dipole-dipole interaction between the magnetic moments is taken into account in an effective through the value of the anisotropy parameter $K$ [6,11] but is not included explicitly in (1). The summation $<i, j>$ in (1) runs over all pairs of nearest neighbor sites. The numerical values of the parameters are taken from [12] and correspond to experimentally measured system consisting of a Pd/Fe overlayer on a Ir(111) substrate [1,13]: $\mu B = 0.093J$, $K = 0.07J$, $D_{ij} = 0.32 \, J$, $J=7$ meV. A metastable skyrmionic state exists in this system as well as the homogeneous FM ground state.

For applications in nano-devices, one possibility is to move skyrmions along on a track. A critical issue is the lifetime of the skyrmion especially when the intrinsic size of the skyrmion is comparable to the width of the track. To describe such a system, we use periodic boundary condition along the track and free boundaries in the perpendicular direction. For the parameters chosen above, the typical size of the skyrmion is several nanometers and the simulation cell has to include at least 50×50 magnetic moments. The system then has 5000 degrees of freedom. Finding the MEP on such a multidimensional energy surface is a non-trivial task but it can be done using geodesic nudged elastic band method [7]. The maximum energy along the MEP, $E^s$, which corresponds to a saddle point on the energy surface, gives the activation energy of the transition as $E^s - E^m$, where $E^m$ is the energy of the initial state minimum. The activation energy determines the exponential dependence of the transition rate on temperature. The pre-exponential factor can be estimated within HTST where a harmonic approximation is made for the energy in the vicinity of the saddle point and minima

$$E_\lambda(\mathbf{q}) = E_\lambda(\mathbf{0}) + \frac{1}{2}\sum_{i=1}^{N}\varepsilon_{\lambda,i}q_{\lambda,i}^2 \qquad (2)$$

Here, the index $\lambda$ corresponds either to the minimum ($\lambda = m$) or the saddle point ($\lambda = sp$). The coordinates q are the deviations from these points along normal mode vectors. $N$ denotes the dimension of the configuration space. The transition state dividing surfaces is taken to be a hyperplane going through the saddle point and having normal vector along the unstable mode, ile. pointing along the eigenvector corresponding to negative eigenvalue of the Hessian. The component of the velocity of system obtained from Landau-Lifshitz equations along the unstable mode at the transition state, can be as a linear combination of the normal mode coordinates $q_i$

$$\dot{q}_{sp,1} = \sum_{i=2}^{N} a_i q_i \qquad (3)$$

The pre-exponential factor, $\nu_0$, within HTST then becomes [3]

$$\nu_0 = \frac{1}{2\pi}\frac{J_{sp}}{J_m}\sqrt{\sum_{i=2}^{N}\frac{a_i^2}{\varepsilon_{sp,i}}}\frac{\prod_{i=1}^{N}\sqrt{\varepsilon_{m,i}}}{\prod_{i=2}^{N}\sqrt{\varepsilon_{sp,i}}} \qquad (4)$$

Here $J_{sp}$ and $J_m$ are Jacobians evaluated at the saddle point and at the minimum, respectively. Eq. (4) is obtained when the harmonic approximation (2) is made for all degrees of freedom in the system. However, there can be degrees of freedom from which the energy of the system does not change, so-called zero-modes. For example, the in-plane movement of a skyrmion as a whole corresponds to two zero-modes. For these degrees of freedom expansion (2) fails and integration over corresponding variables has to be performed separately. In this case, the variable $N$ in (4) should be taken to be the number of non-zero modes and expression (4) for the pre-exponential factor contains an addition multiplicative factor

$$\frac{V_{sp}}{V_m}(2\pi k_B T)^{\frac{n_m - n_{sp}}{2}}, \qquad (5)$$

where $V_{sp}$ and $V_m$ are volumes, and $n_{sp}$ and , $n_m$ the numbers of zero modes at the saddle point and at the initial state minimum. If $n_m \neq n_{sp}$, the attempt frequency will depend on temperature as given by eqn. (5). Knowledge of the activation energy and pre-exponential factor makes it possible to estimate the lifetime of the magnetic states at any value of the temperature for which HTST is applicable as $\tau = \nu_0^{-1}\exp[(E^s - E^m)/k_B T]$.

### 3. Skyrmion on a track

For a skyrmion on a track, as in a racetrack memory device, it is important to assess the influence of the boundaries of the track on the stability of skyrmion. When the intrinsic the size of the skyrmion is comparable to the width of the track, the activation energy for skyrmion annihilation is smaller and, correspondingly, the activation energy for nucleation is higher than for a skyrmion in an extended domain [8]. For very narrow tracks the skyrmion is not even metastable. There are two different mechanisms for skyrmion nucleation and annihilation in a track, either in the interior or through the boundary of the track. The two mechanisms were investigated for wide tracks in refs. [9,10] and as function of the track width for a skyrmion in a magnetic field of B=3.75 T in ref. [8]. The activation energy for skyrmion annihilation in the interior of the track was found to be higher than escape through the boundary in refs. [8,9], whereas in ref. [10] a crossover with respect to the external magnetic field was identified: at a large enough magnetic field the activation energy for boundary escape became higher than for annihilation in the interior of the track.

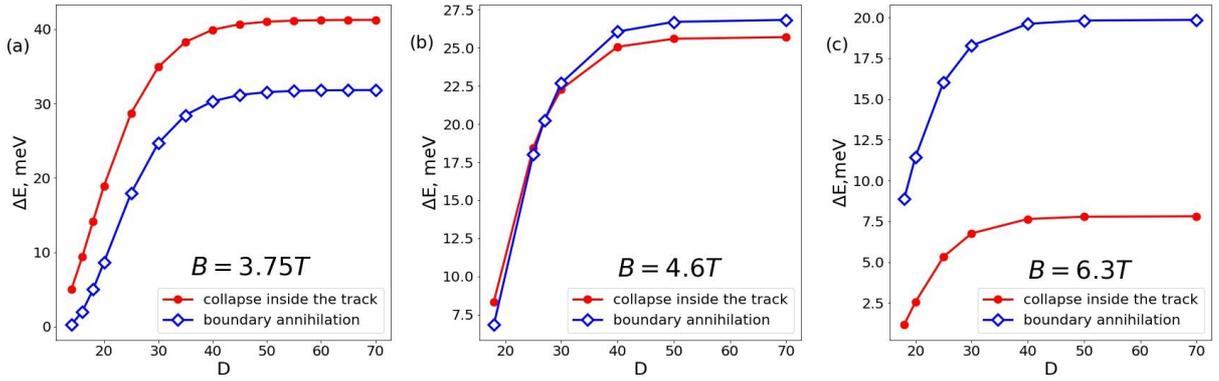

Fig 1 Variation of the activation energy for annihilation of a skyrmion as a function of the width of the track, D, for three different values of the magnetic field
B = 3.75 T in (a), B = 4.6 T in (b), and B = 6.3 T in (c). Solid red symbols correspond to annihilation in the interior of the track, whereas open blue symbols corresponds to escape through the boundary.

The activation energy for skyrmion annihilation in the interior of the track and escape through the boundary is given as a function of the track width D in Fig. 1 for three different values of the magnetic field. For a field of B = 3.75 T, as in ref. [8], escape through the boundary has lower activation energy for all values of D. In large magnetic field of B=6.3 T, the opposite order in activation energy is obtained, similar to what has been presented in [10]. A crossover takes place at in a field of B = 4.6 T, where for a narrow track the activation energy is lower for escape through the boundary, but for a wide track (D>27) it is lower for annihilation in the interior of the track.

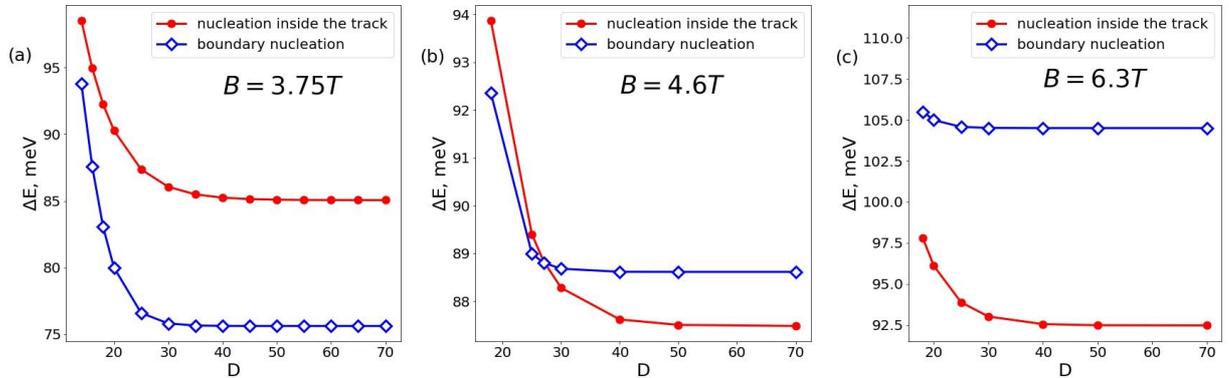

Fig 2. Variation of the activation energy for nucleation of a skyrmion as a function of the width of the track, D, for three different values of the magnetic field
B = 3.75 T in (a), B = 4.6 T in (b) and B = 6.3 T in (c). Solid red symbols correspond to nucleation in the interior of the track, whereas open blue symbols correspond to nucleation at the boundary.

The variation of the activation energy for nucleation of a skyrmion with the width of the track is opposite to that of annihilation, as shown in Fig. 2. Again, a crossover is observed for a field of B = 4.6 T, where for a wide track nucleation is lower in the interior of the track (D>27) but is lower for formation at the boundary for a narrow track (D<27).

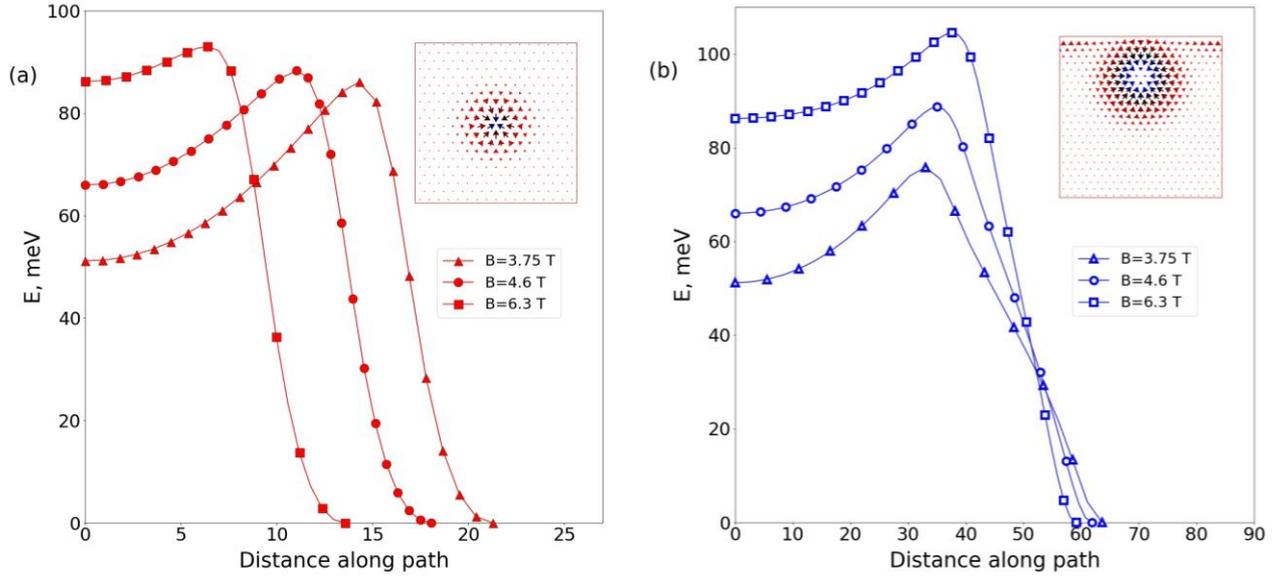

Fig. 3. Minimum energy paths for the transition from the skyrmion state to the ferromagnetic state in the interior (a) and through the boundary (b) of a track. Insets show saddle point configurations of the magnetic vectorsat. The width of the track is 30 atomic rows.

To clarify the reason for the crossover shown in Fig. 3, the MEPs for skyrmion annihilation inside the track (a) and via the boundary of the track are shown for the three different values of the magnetic field. When the magnetic field is increased, the energy of the skyrmion state as compared to the FM ground state increases as well as the energy of the transition state, and the skyrmion becomes smaller. For annihilation in the interior of the track, the saddle point energy grows slower with the magnetic field than for escape through the boundary. As a result, the activation energy for escape through the boundary becomes larger than for annihilation in the interior of the track when the field reaches a certain value. The evolution of the magnetic configuration along the MEP is shown in the supplementary information for annihilation in the interior of the track and in for escape through the boundary [14].

The lifetime of a skyrmion is determined not only by the activation energy but also by the attempt frequency. For annihilation in the interior of the track, both the skyrmion state and the saddle point have two zero-modes which correspond to in-plane translation of the magnetic configuration. Therefore, according to (5), the attempt frequency does not depend on temperature. and the values obtained for the attempt frequency are $2.4 \cdot 10^{12}$ s$^{-1}$, $2.1 \cdot 10^{12}$ s$^{-1}$ and $1.2 \cdot 10^{12}$ s$^{-1}$ for the mangetic field of 3.75 T, 4.6 T and 6.3 T, respectively. For escape through the boundary, there is only one zero-mode at the saddle point corresponding to translation along the border. In this case the attempt frequency is proportional to $(k_B T)^{1/2}$. For T=10 K the values obtained are $2.0 \cdot 10^{12}$ s$^{-1}$, $0.9 \cdot 10^{12}$ s$^{-1}$ and $0.4 \cdot 10^{12}$ s$^{-1}$ for the three values of the magnetic field.

### 4. Skyrmion at a nonmagnetic impurity

The interaction of a skyrmion with defects can be an important consideration for the stability and dynamics. Experiments have indicated that atomic defects can act as nucleation and pinning sites for skyrmions [13]. There is a repulsive interaction with non-magnetic defects but if this energy is overcome the skyrmion tends to be pinned by such defects [15,16]. It is important to characterize the energy surface for a skyrmion near an impurity, and assess the activation energy for the attachement and dissociation as well as the effect of the impurity on the lifetime of the skyrmion.

Non-magnetic defects were modeled by empty sites of the triangular lattice. Figure 4 show the MEP for attachment and subsequent annihilation at a defect corresponding to three non-magnetic impurity atoms. The initial state corresponds to a skyrmion placed far from the defect. After overcoming a small energy barrier, a lower energy minimum is obtained corresponding to a stable state where the defect is near the border of the skyrmion, where the magnetic moments have a large in-plane component. This is in agreement with experimental observations [15]. The saddle point on the energy surface between these two states determines the activation energy for pinning and dissociation of the skyrmion from the defect.

The second part of the MEP shown in Fig.4 describes the annihilation of the skyrmion at the defect. Comparison of the activation energy for annihilation and nucleation with the data depicted in Figs. 1a and 2a for large values of D shows that the defect strongly decreases the values of the activation energy. Therefore, the calculated MEP shows that such defects decrease the lifetime of skyrmions and increase the rate of magnetic transitions. This is in agreement with experimental observations [13].

An interesting possibility is to make use of the pinning of a skyrmion by non-magnetic defects to create a preferred path for the motion of a skyrmion as it hops from one defect to another. This would be an alternative to a track for directing its motion.

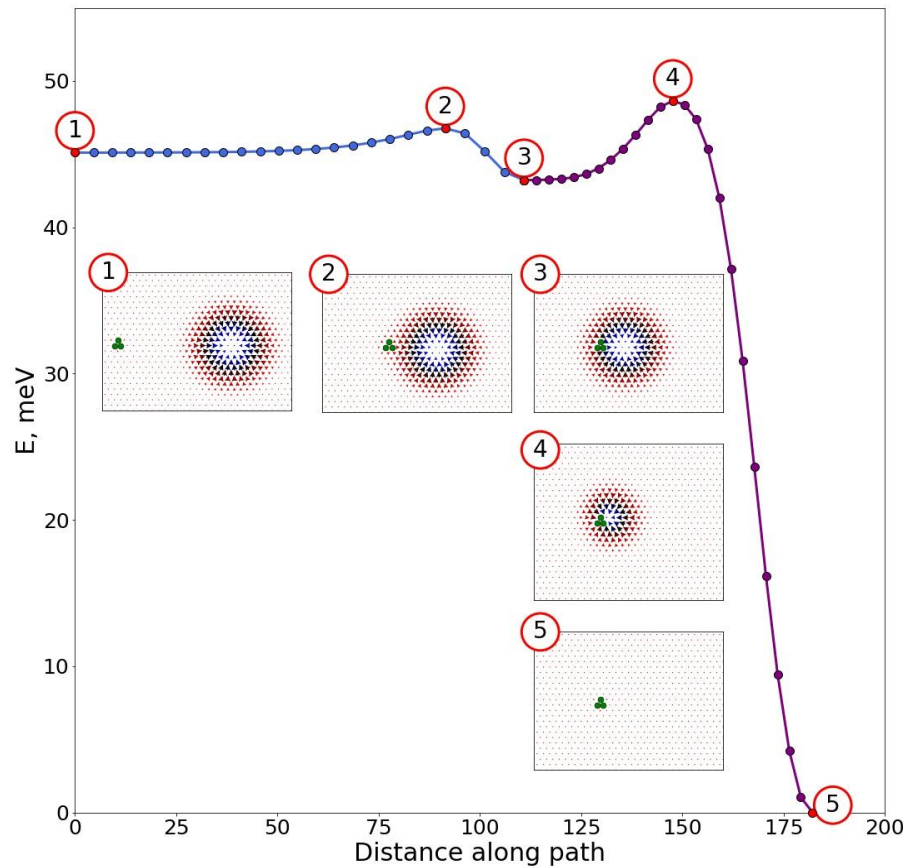

Fig. 4 Minimum energy path for the attachment of a skyrmion to a non-magnetic impurity consisting of three non-magnetic atoms (the initial state, saddle point and final state marked with 1, 2 and 3) and for the subsequent annihilation (saddle point and final state marked with 4 and 5). Insets show the corresponding orientation of the magnetic moments. The filled circles on the curve show location of images in the GNEB calculation. .

**Acknowledgments**

This work was supported by the Icelandic Research Fund and the Academy of Finland (grant 278260).